\documentclass[conference]{IEEEtran}
\IEEEoverridecommandlockouts
\usepackage{cite}
\usepackage{amsmath,amssymb,amsfonts}
\usepackage{graphicx}
\usepackage{textcomp}
\usepackage{xcolor}
\usepackage{comment}
\usepackage{inconsolata} 
\usepackage{enumitem}
\usepackage[hidelinks]{hyperref} 
\usepackage[T1]{fontenc}

\makeatletter
\def\mdseries@tt{m}
\makeatother
\usepackage[frozencache=true]{minted}
\fvset{linenos, fontsize=\scriptsize, frame=lines, numbersep=1pt, xleftmargin=.75em}

\usepackage[font=scriptsize]{caption}
\usepackage[font=scriptsize]{subcaption}
\usepackage{listings}
\DeclareCaptionSubType{listing} 

\usepackage[per-mode=symbol]{siunitx}
\DeclareSIUnit \bit {bit}
\DeclareSIUnit \byte {Byte}
\DeclareSIUnit \cycle {cycle}
\DeclareSIUnit \cycles {cycles}
\DeclareSIUnit \hz {Hz}
\DeclareSIUnit \op {Op}
\DeclareSIUnit \operand {operand}
\DeclareSIUnit \operands {operands}
\DeclareSIUnit \transfer {T}
\DeclareSIUnit \cell {cell}

\newcommand{\code}[1]{\texttt{#1}}
\newcommand{\hlslib}{\code{hlslib}}

\newcommand{\figureref}[1]{Figure~\ref{fig:#1}}

\newcommand{\listingref}[1]{Listing~\ref{lst:#1}}

\newcommand{\sectionref}[1]{Section~\ref{sec:#1}}

\newcommand{\secsymbol}{{\S}}

\begin{document}

\title{\resizebox{\textwidth}{!}{\hlslib: Software Engineering for
Hardware Design}}

\author{%
\IEEEauthorblockN{Johannes de Fine Licht}
\IEEEauthorblockA{\textit{ETH Zurich}\\
definelicht@inf.ethz.ch}
\and
\IEEEauthorblockN{Torsten Hoefler}
\IEEEauthorblockA{\textit{ETH Zurich}\\
htor@inf.ethz.ch}%
}

\maketitle

\begin{abstract}
High-level synthesis (HLS) tools have brought FPGA development into the
mainstream, by allowing programmers to design architectures using familiar
languages such as C, C++, and OpenCL\@. 
While the move to these languages has brought significant benefits, many aspects
of traditional software engineering are still unsupported, or not exploited by
developers in practice.
Furthermore, designing reconfigurable architectures requires support for
hardware constructs, such as FIFOs and shift registers, that are not native to
CPU-oriented languages. 
To address this gap, we have developed \hlslib{}, a collection of software
tools, plug-in hardware modules, and code samples, designed to enhance the
productivity of HLS developers.
The goal of \hlslib{} is two-fold: first, create a community-driven arena
of bleeding edge development, which can move quicker, and provides more powerful
abstractions than what is provided by vendors; and second, collect a wide range
of example codes, both minimal proofs of concept, and larger, real-world
applications, that can be reused directly or inspire other work. 
\hlslib{} is offered as an open source library, containing CMake files, C++
headers, convenience scripts, and examples codes, and is receptive to any
contribution that can benefit HLS developers, through general functionality or
examples.
\end{abstract}


\section{State-of-the-art}

Developing for FPGAs gives programmers an empty slate to lay out a custom
architecture that implements a target application. This is the biggest strength
of reconfigurable hardware, but also its biggest weakness: achieving performance
that is competitive with software -- in particular when comparing to non-naive
GPU implementations -- often requires a tremendous amount of effort.
%
Although the productivity of developing for FPGAs has improved significantly
with widespread adoption of HLS~\cite{xilinx_hls}, working with these tools is
notorious for being a less-than-smooth experience. There are multiple reasons
for this.
The imperative languages primarily used by HLS tools, namely C, C++, and
OpenCL, were not designed with hardware development in mind, and the resulting
opaque mapping to hardware frustrates both software developers (who cannot
implement code in the way they are used to), and hardware developers (who
struggle to achieve the exact architecture that they have in mind).
%
%
Furthermore, the placement and routing process that maps a synthesized
architectures to the FPGA chip is a time consuming process, where bigger
designs can take up to a full day to compile, which inhibits iterative
debugging and development.  
Finally, because of the additional layer of abstraction added by HLS, tracking
problems in the final architecture back to the origin in the HLS code can be
near impossible, which sometimes results in debugging and optimization
degenerating into a trial-and-error process.
%

\begin{figure}[h]
  \centering
  \includegraphics[width=.9\columnwidth]{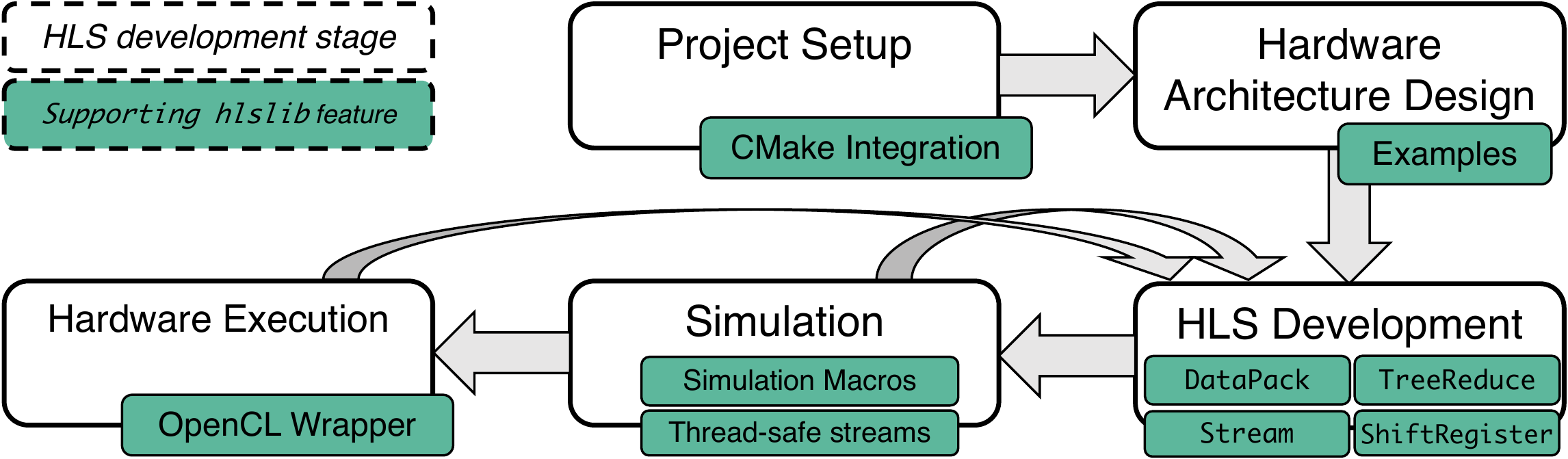}
  \caption{Stages of HLS development and the supporting \hlslib{} features.}
  \label{fig:workflow}
\end{figure}

We introduce \hlslib{}\footnote{Available at
\url{https://github.com/definelicht/hlslib}}, an open source collection of
tools, modules, scripts, and examples, with the overarching goal of improving
the quality of life of HLS developers. An overview of some \hlslib{} features
and which stage of development benefits from them is given in
\figureref{workflow}. 
While \hlslib{} cannot hope to solve all the issues of HLS development, we hope
to smoothen as many steps of the process as possible in a external library, and
encourage good practices inspired by traditional software engineering.
The following sections give an overview of the functionality offered by
\hlslib{} as of writing this work, but the library is continuously developed to
provide new features and to support newer versions and functionality of the two
major vendor tools, Xilinx' Vivado~HLS~\cite{autopilot}, and Intel's OpenCL SDK
for FPGAs~\cite{intel_fpga_opencl}. 

\section{Improving the HLS Workflow}

\subsection{CMake Integration}
\label{sec:cmake}

Many published HLS projects, including example codes by Xilinx and Intel, rely
on manually written GNU makefiles. This method offers poor portability, and
does not allow projects to be configured without modifying the makefile or
source code.
In software development, CMake is a widespread tool used to configure and build
C/C++ projects. Users can set project parameters during configuration, as
compilation is performed out-of-source, and dependencies are automatically
located on the host system in a portable fashion. 

\hlslib{} provides supports for FPGA projects in CMake, allow separation of
source code and configuration through the \texttt{FindSDAccel.cmake} and
\texttt{FindIntelFPGAOpenCL.cmake} scripts, required to locate and expose the
Xilinx and Intel FPGA ecosystems, respectively. Users gain access to the HLS
binaries, as well as compiler flags, header files, and library files required
to build the OpenCL host code.
Historically, the workflow for building and running FPGA codes with commercial
HLS tools has been continuously changing throughout their development.  By
offloading the responsibility of setting up the HLS environment to \hlslib{},
projects become robust to changes in the setup provided by the vendors. 

An example snippet for a \texttt{CMakeLists.txt} using \hlslib{} to build an
SDAccel project with a top-level function ``\texttt{Top}'' is given below, where
hardware targets are added with custom targets, using the binaries exposed by
the find-scripts:
\begin{listing}[h]
  \begin{minted}{CMake}
set(CMAKE_MODULE_PATH hlslib/cmake)
find_package(SDAccel REQUIRED)
include_directories(${SDAccel_INCLUDE_DIRS})
add_executable(MyHostCode MyHostCode.cpp)
target_link_libraries(MyHostCode ${SDAccel_LIBRARIES})
add_custom_target(compile_hardware COMMAND ${SDAccel_XOCC}
                  --kernel Top -c -t hw Kernel.cpp -o Kernel.xo)
add_custom_target(link_hardware COMMAND ${SDAccel_XOCC}
                  --kernel Top -l -t hw Kernel.xo -o Kernel.xclbin)
  \end{minted}
  \vspace{-0.5em}
  \caption{Creating custom FPGA targets with \hlslib{} CMake support.}
  \label{lst:cmake}
\end{listing}

\noindent Examples of the full flow with all relevant files are included in the
\hlslib{} repository, for both Xilinx and Intel OpenCL ecosystems.

\subsection{Portable OpenCL Host Code}
\label{sec:opencl}

OpenCL was originally developed for GPUs, and thus follows the GPU model of
creating computational kernels, transferring data in bulk between host and
device memories, and launching kernels synchronously.  
OpenCL is exposed as a host-side interface by both Intel and Xilinx for
launching computational kernels and interacting with device DRAM.

Intel and Xilinx have taken slightly different approaches to adapting OpenCL to
FPGAs. In order to enable a fully unified interface, \hlslib{} provides an
OpenCL wrapper that hides subtle differences between vendors, such as single
command queue (Xilinx) versus one-queue-per-kernel (Intel), and extended memory
pointer (Xilinx) versus a simple memory flag (Intel) for specifying FPGA memory
banks.
An example of a basic \hlslib{} OpenCL host program is given in
\listingref{opencl_host}, which is valid code for both the Intel and Xilinx
ecosystems (example file name uses the Xilinx \texttt{.xclbin} suffix).

\begin{listing}[h]
  \begin{minted}{C++}
using namespace hlslib::ocl;
Context context; // Sets up the vendor OpenCL runtime
auto program = context.MakeProgram("KernelFile.xclbin");
std::vector<float> input_host(N, 5), output_host(N);   
auto input_device = context.MakeBuffer<float, Access::read>(
    MemoryBank::bank0, input_host.cbegin(), input_end.cend());
auto output_device = context.MakeBuffer<float, Access::write>(
    MemoryBank::bank1, N);
auto kernel = program.MakeKernel("Kernel", in_device, out_device, N);
kernel.ExecuteTask(); // Synchronous kernel execution
output_device.CopyToHost(output_host.begin());
  \end{minted}
  \vspace{-0.5em}
  \caption{Portable OpenCL host program implemented with the \hlslib{} wrapper.}
  \label{lst:opencl_host}
\end{listing}


\subsection{Emulating Multiple Processing Elements in Software}
\label{sec:simulation}
\label{sec:emulation}

Accurately emulating the semantics of multiple concurrent processing elements
(PEs) executing in hardware is critical to the testing process, as multiple PEs
are vital to any high performance architecture. PEs typically communicate via
blocking channels, implying synchronization points between them. Emulating
concurrent PEs thus requires a multi-threaded environment with thread-safe
constructs.

In the Intel OpenCL ecosystem, PEs are expressed as OpenCL kernels that are
launched separately from the host code, and communication channels are expressed
as global objects that are accessed within the kernel codes. When running
emulation in software, PEs are thus launched as concurrent threads by the
runtime. On the other hand, Xilinx HLS instantiates PEs from functions or loops
``called'' in a scope annotated with the \texttt{DATAFLOW} pragma. While this
allows expressing communication between kernels with multiple PEs in a more
explicit fashion, it also means that the behavior of executing the code when
compiled as C++ code can differ significantly from its behavior when built for
hardware. An example of this is shown in \listingref{dataflow_divergence}, when
\texttt{mem0} and \texttt{mem1} are passed as pointers \emph{to the same
address}:

\begin{listing}[h]
  \begin{minipage}[t]{.52\columnwidth}
    \begin{minted}{C++}
void Top(int const *mem0,
         int *mem1) {
  #pragma HLS DATAFLOW
  hlslib::Stream<int> s0, s1;
  Read(mem0, s0);   // Sequential
  Compute(s0, s1);  // in software,
  Write(s1, mem1);  // parallel
}                   // in hardware. 

void Compute(hlslib::Stream &s0,
             hlslib::Stream &s1) {
  for (int t = 0; t < T; ++t)
    for (int i = 0; i < N; ++i) {
      #pragma HLS PIPELINE
      int read = s0.Pop();
      int res = /* do compute */;
      s1.Push(res);
    }
}
    \end{minted}
  \end{minipage}\hfill%
  \begin{minipage}[t]{.48\columnwidth}
    \begin{minted}[stripnl=false]{C++}
void Read(int const *mem0,
          hlslib::Stream &s) {
  for (int t = 0; t < T; ++t)
    for (int i = 0; i < N; ++i)
      #pragma HLS PIPELINE
      s.Push(mem0[i]);  }

void Write(hlslib::Stream &s,
           int *mem1) {
  for (int t = 0; t < T; ++t)
    for (int i = 0; i < N; ++i)
      #pragma HLS PIPELINE
      mem1[i] = s.Pop(); }
    \end{minted}
    \centering
    \includegraphics[width=.9\textwidth]{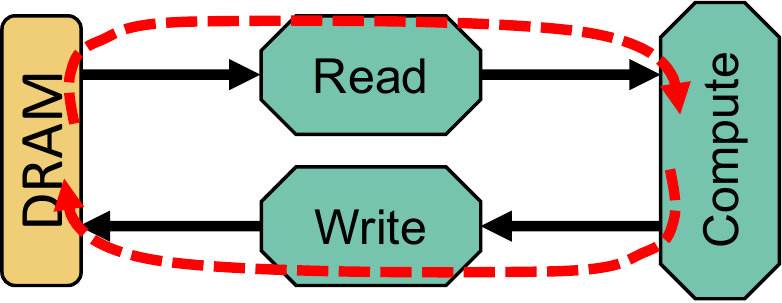}
  \end{minipage}
  \begin{minipage}{\columnwidth}
  \end{minipage}
  \vspace{-0.5em}
  \caption{Software and hardware behavior is different for cyclic dataflow.}
  \label{lst:dataflow_divergence}
\end{listing}

Programs with cyclic dataflow between PEs are not officially supported by
Xilinx, but will compile and run in practice, albeit without any guarantees of
correctness.  It is often desirable to write such programs for high-performance
implementations of iterative algorithms, such as iterative stencil computations,
where the same DRAM memory addresses are read and written multiple times during
execution. In such a scenario, a program like \listingref{dataflow_divergence}
will exhibit different behavior in software and hardware:
\begin{itemize}[leftmargin=*]
  \item In software, \texttt{Read} will execute all $T \cdot N$ iterations
  before \texttt{Compute} is called, which will execute all $T \cdot N$
  iterations before \texttt{Write} is called. Assuming that the streams
  \texttt{s0} and \texttt{s1} are unbounded in emulation, each iteration $t$
  will perform exactly the same computation.
  \item In hardware, \texttt{Read}, \texttt{Compute}, and \texttt{Write} will
  run concurrently, and \texttt{s0} and \texttt{s1} will be bounded with size
  $1$. The PEs will thus stay synchronized. Each iteration $t$ of \texttt{Read}
  will read values written by the previous iteration of \texttt{Write}, assuming
  $N$ is significantly larger than the pipeline depth.
\end{itemize}
In the best case, programs will crash or not terminate in software, when
feedback happens directly between PEs, where there is a cycle in the channels
interconnecting them.  In the worst case, programs like
\listingref{dataflow_divergence} will produce different results in software and
hardware, because the feedback dependency on \texttt{mem} is not enforced.  

To correctly emulate kernels with multiple PEs and feedback dependencies,
\hlslib{} provides a set of thin wrapper macros that mitigate the difference
between the compiled C++ and the hardware generated HLS, that can be used in
conjunction with \texttt{hlslib::Stream} wrapper objects (see
\sectionref{streams}) to run PEs concurrently synchronously. Programs only need
to wrap every function call in a \texttt{DATAFLOW} section in an
\hlslib{}-defined macro, as shown in \listingref{simulation_macros}, which is a
modified version of the top-level function from
\listingref{dataflow_divergence}.

\begin{listing}[h]
  \begin{minted}{C++}  
void Top(int const *mem0, int *mem1) {
  #pragma HLS DATAFLOW
  hlslib::Stream<int> s0, s1; // hlslib streams are thread-safe
  HLSLIB_DATAFLOW_INIT();
  HLSLIB_DATAFLOW_FUNCTION(Read, mem0, s0);  // In simulation mode,
  HLSLIB_DATAFLOW_FUNCTION(Compute, s0, s1); // each call launches 
  HLSLIB_DATAFLOW_FUNCTION(Write, s1, mem1); // a separate C++ thread 
  HLSLIB_DATAFLOW_FINALIZE(); // Joins C++ threads
}                 
  \end{minted}
  \vspace{-0.5em}
  \caption{PEs in \texttt{DATAFLOW} section annotated to emulate hardware
  behavior.}
  \label{lst:simulation_macros}
\end{listing}

\vspace{-0.5em}
\noindent Behind the scenes, each \texttt{HLSLIB\_DATAFLOW\_FUNCTION} macro
chooses between two kinds of behavior, depending on the compilation mode:
\begin{itemize}[leftmargin=*]
  \item In hardware, all annotated functions are simply inlined, resulting in
  code identical to \listingref{dataflow_divergence}.
  \item In software, each function is executed in a newly launched C++ thread.
  When \texttt{HLSLIB\_DATAFLOW\_FINALIZE} is called, \hlslib{} will wait on
  each of the launched threads, returning when all PEs have terminated.
\end{itemize}
The software behavior means that PEs cannot run ahead of others more than what
is allowed by the depth of the channels between them, which also allows
debugging deadlocks due to channel sizes (i.e., depth of the FIFOs implementing
them in hardware). When a \texttt{Pop} or \texttt{Push} from/to a channel has
waited for a configurable amount of time without receiving data, \hlslib{} will
print a warning with the channel name and operation, enabling easier debugging
of deadlocks.

\section{Object-Oriented Hardware Design}

Classes can provide excellent encapsulation for hardware concepts, combining
data and functionality in the spirit of object-oriented programming, but also
allow specializing classes with C++ templates allows parameters to be specified
at compile-time, when this is necessary for generating hardware. \hlslib{} uses
classes both in the object-oriented sense, and by exploiting template
metaprogramming, to fill various gaps in hardware development.

\subsection{Streams/Channels}
\label{sec:stream}
\label{sec:streams}
\label{sec:channels}

Channel objects are ubiquitous in HLS programming, either as communication
primitives between processing elements, or as buffers with FIFO semantics.
Whereas channels in Intel OpenCL are global objects, channels are created in
Vivado~HLS as templated \texttt{hls::stream} objects, and act not only as
inter-PE connections, but also as buffers internal to a single module when a
queue-like buffering pattern is sufficient.

\hlslib{} extends the Vivado~HLS built-in \texttt{hls::stream} class in the
\texttt{hlslib::Stream} class, which adds a number of additional features and
streamlines the interface. Most notably, \hlslib{} streams is thread-safe, and
supports the features offered by the \hlslib{} multiple-PE simulation
functionality (example usage shown in
Listings~\ref{lst:dataflow_divergence}~and~\ref{lst:simulation_macros}).
Furthermore, streams are bounded by default, like the hardware implement they
represent. If no argument is specified, the default Vivado~HLS implementation is
used, which is a ping-pong buffer. Any other depth will implement a FIFO using a
resource suggested by the tool, or specified by an optional template argument
(e.g., SRL, LUTRAM, BRAM, or UltraRAM). 

\subsection{Wide Data Buses and Vectorization}
\label{sec:datapack}

Instantiating wide data paths in HLS is necessary to exploit memory
bandwidth~\cite{hls_transformations}, and to achieve parallel architectures
through vectorization. In practice, this is typically done either by unrolling
loops and relying on the tool to infer wide data accesses, or by using types
that explicitly specify the vector size, such as OpenCL vector types for Intel
OpenCL, or \texttt{ap\_uint} for Vivado~HLS\@. OpenCL vector types only expose a
small, limited set of types and vector lengths, and \texttt{ap\_uint} requires
tedious and error-prone casting to implement vector types in hardware.

\hlslib{} provides the templated \texttt{DataPack} class for Vivado~HLS, which
exposes a versatile interface for implementing wide buses, registers, memory
interfaces, and computations that consist of multiple data elements. Unlike
\texttt{ap\_uint}, \texttt{DataPack} is typed, allowing native indexing of
elements for both reading and writing, supports element-wise operations (shown
in \listingref{datapack}), and convenience functions for concerting to and from
C-style arrays and \texttt{ap\_uint} types.

\begin{listing}[h]
  \begin{minted}{C++}
using hlslib::DataPack; // Import class into namespace
DataPack<float, 4> Direction(DataPack<float, 4> &a,
                             DataPack<float, 4> &b) {
  auto d = b - a; // Vector operations 
  auto len = c[0] + c[1] + c[2] + c[3]; // Indexing
  return 1/len * d; // Element-wise operations 
}
  \end{minted}
  \vspace{-0.5em}
  \caption{Overview of \texttt{hlslib::DataPack} functionality.}
  \label{lst:datapack}
\end{listing}

\vspace{-0.5em}
When used as the data type for pointer or stream arguments, \texttt{DataPack}
enforces bus widths corresponding the byte width specified by the data type
vector size. If used consistently, simply changing the width of a centrally
defined \texttt{DataPack}-based type will be sufficient to adjust registers,
buses, buffers, and interfaces throughout an HLS code. 

\subsection{Shift Registers with Parallel Access}
\label{sec:shift_registers}

A common pattern for FPGA algorithms~\cite{zohouri_opencl} is to buffer
elements streamed in for a \emph{constant} number of cycles, thus ``delaying''
them for future iterations (e.g., to be used as a different element of a
\emph{sliding window} in a stencil computation~\cite{sliding_window,
zohouri_stencil}).  This is similar to a FIFO buffer, with the added constraint
that elements pushed and popped are at a constant distance (e.g., for a buffer
of size 4, an element pushed can only be accessed again when it comes out at
the end, i.e., after 4 additional pushes). We will refer to these types of
buffers as \emph{shift registers} according to the Intel FPGA nomenclature,
although it also common to implement these in BRAM/M20K on-chip memory.  We
assume that shift registers have a single input, but can have multiple parallel
outputs (known as ``taps'').

In Intel OpenCL, shift registers are inferred as a pattern when an unrolled
loop shifts an array by a constant offset every cycle of a pipelined section,
and the remainder of the section only accesses the array using constant
indices. 
%
The compiler can then infer the distance between each tap, allowing it to
instantiate separate buffers in hardware between them, effectively partitioning
the single array into multiple smaller buffers. Vivado~HLS, on the other hand,
does not recognize this as a high-level pattern (as of writing this work), and
will textually unroll the shifting loop in the preprocessor and analyze the
unrolled code, which does not scale with large shift registers.

We express the parallel shift register abstraction as a templated class in
\hlslib{}, transparently managing buffers between each tap. Unlike the Intel
ecosystem, \hlslib{} shift registers are explicitly instantiated by the
programmer (as opposed to relying on pattern detection), and enforce constant
offset access at compile-time, while providing the full abstraction to the
Vivado~HLS ecosystem, which otherwise requires this pattern to be implemented
manually. An implementation of a 4-point 2D stencil code based on an \hlslib{}
shift register is shown in \listingref{shift_register}.

\begin{listing}[h]
  \centering
  \includegraphics[height=5em]{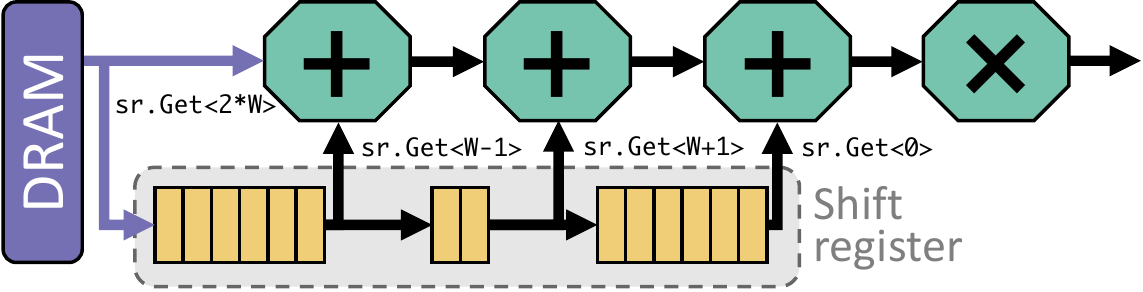}
  \begin{minted}{C++}
void Stencil(hlslib::Stream<float> &in, hlslib::Stream<float> &out) {
  // Explicitly declare taps as template arguments
  hlslib::ShiftRegister<float, 0, W - 1, W + 1, 2 * W> sr;
  // H and W are compile-time constants
  for (int i = 0; i < H; ++i) {
    for (int j = 0; j < W; ++j) {
      #pragma HLS PIPELINE
      sr.Shift(in.Pop()); // Push new element and shift buffer
      if (i >= 2 && j >= 1 && j < W - 1) { // Ignore boundary 
        // Specify tap to access using compile-time indices
        float res = 0.25 * (sr.Get<2 * W>() + sr.Get<W - 1>() +
                            sr.Get<W + 1>() + sr.Get<0>());
        out.Push(res);
} } } }
  \end{minted}
  \vspace{-0.5em}
  \caption{Explicit shift register abstraction provided by \hlslib{}.}
  \label{lst:shift_register}
\end{listing}

Variadic template arguments are used to instantiate taps, where the distance
between each consecutive index is used to compute the respective buffer size
(as a result, indices must be specified in ascending order).



\subsection{Tree Reduction with Functors}
\label{sec:tree_reduction}
\label{sec:functor}
\label{sec:functors}

To perform a fully pipelined reduction of an array of elements for an
associative operator, it is common to implement the reduction as a balanced
binary tree to minimize latency and resource utilization. 
Implementing reduction trees in an imperative language requires the compiler to
recognize unrolled loops that accumulate into a single variable, and requires
explicitly allowing the compiler to reorder non-associative operations, such as
floating point addition.

To guarantee that a reduction is performed as a balanced binary tree, \hlslib{}
provides the \code{TreeReduce} templated function, which uses variadic templates
to explicitly instantiate the full tree in hardware. The template supports any
type, array size, and binary operator. An example is shown below:
\begin{listing}[h]
  \begin{minted}{C++}
using Vec = DataPack<float, 8>;
void Reduce(hlslib::Stream<Vec> &in, hlslib::Stream<Vec> &out) {
  for (int i = 0; i < 1024; ++i) {
    #pragma HLS PIPELINE
    auto v = in.Pop();
    auto r = hlslib::TreeReduce<float, hlslib::op::Add<float>, 8>(v);
    out.Push(r);
} }
  \end{minted}
  \vspace{-0.5em}
  \caption{Explicit balanced tree reduction of an array.}
  \label{lst:tree_reduce}
\end{listing}

\noindent\hlslib{} supports a set of common binary operators by default, but
custom operators can be implemented with a functor struct that defines the
\code{Apply} binary function and an \code{identity} for the operator. These
functors are conveniently expressible using C++ templated classes.

\section{Open Source Development with \hlslib{}}

All the features described in this work were tested to meet the demands of
concrete HLS codes. The repository holds additional niche features left out
here, as well as a compilation of examples testing and demonstrating various
concepts.

We maintain a list of projects leveraging \hlslib{} on the repository page, and
noteworthy examples include:
\label{sec:dace}
\label{sec:dapp}
the \emph{Data Centric Parallel Programming} (DaCe) project~\cite{dapp}, a
data-centric optimization framework targeting a multitude of backends,
including code generation for both Xilinx and Intel FPGAs;
%
\label{sec:smi}
and the reference implementation of the \emph{Streaming Message Interface}
(SMI)~\cite{smi}, a distributed memory inter-FPGA communication model
specification unifying message passing with the streaming model of pipelined
HLS codes.


\vspace{0.5em}


As the field develops, \hlslib{} will continue to evolve and adapt to new tool
features and new ideas for how to close the productivity gap.  However, to truly
accelerate HLS development, the field must see many open source efforts, with
active \emph{exchange} of knowledge and pooling of developer effort, so that
hardware design can reap the benefits of open source development that we know
from the software domain. 

\bibliographystyle{ieeetr}
\bibliography{H2RC19}

\end{document}